\documentclass[twocolumn,english]{revtex4}
\usepackage[T1]{fontenc}
\usepackage[latin9]{inputenc}
\usepackage{amsmath}
\usepackage{amssymb}
\usepackage{graphicx}

\makeatletter

\providecommand{\tabularnewline}{\\}

\@ifundefined{textcolor}{}
{%
 \definecolor{BLACK}{gray}{0}
 \definecolor{WHITE}{gray}{1}
 \definecolor{RED}{rgb}{1,0,0}
 \definecolor{GREEN}{rgb}{0,1,0}
 \definecolor{BLUE}{rgb}{0,0,1}
 \definecolor{CYAN}{cmyk}{1,0,0,0}
 \definecolor{MAGENTA}{cmyk}{0,1,0,0}
 \definecolor{YELLOW}{cmyk}{0,0,1,0}
}

\usepackage{babel}

\newcommand{\ket}[1]{|#1 \rangle}

\usepackage{babel}

\usepackage{babel}

\usepackage{babel}

\usepackage{babel}

\makeatother

\usepackage{babel}
\begin{document}

\title{Scalable universal holonomic quantum computation realized with an
adiabatic quantum data bus and potential implementation using superconducting
flux qubits}

\author{Nicholas Chancellor and Stephan Haas}

\address{Department of Physics and Astronomy and Center for Quantum Information
Science \& Technology, University of Southern California, Los Angeles,
California 90089-0484, USA}
\begin{abstract}
In this paper we examine the use of an adiabatic quantum data transfer
protocol to build a universal quantum computer. Single qubit gates
are realized by using a bus protocol to transfer qubits of information
down a spin chain with a unitary twist. This twist arises from altered
couplings on the chain corresponding to unitary rotations performed
on one region of the chain. We show how a controlled NOT gate can
be realized by using a control qubit with Ising type coupling. The
method discussed here can be extended to non-adiabatic quantum bus
protocols. We also examine the potential of realizing such a quantum
computer by using superconducting flux qubits. 
\end{abstract}
\maketitle

\section*{Introduction}

\begin{figure}
\includegraphics[scale=0.4]{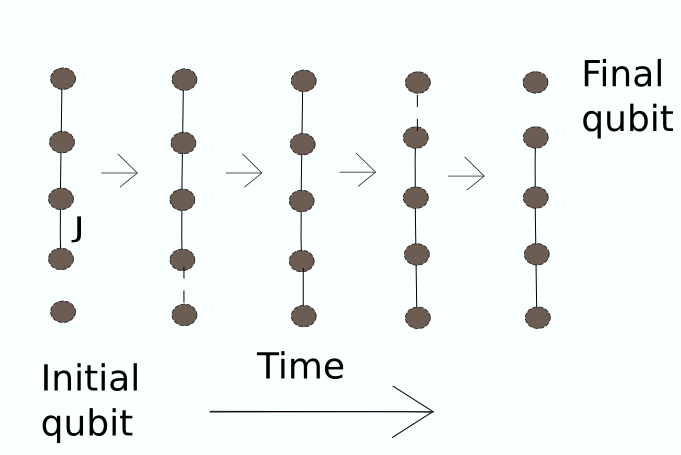}

\caption{\label{fig:transport cartoon} Cartoon of an adiabatic quantum bus
protocol for the Heisenberg spin chain \cite{Chancellor2012}. A spin
with the encoded qubit is connected to one end of an even length antiferromagnetic
chain. Afterwards, the spin on the opposite end is removed adiabatically.
As long as the chain interactions are predominately antiferromagnetic
and the adiabatic theorem is satisfied the qubit will be transferred.}
\end{figure}

It has recently been demonstrated how an open-ended antiferromagnetic
Heisenberg spin chain can be used as an adiabatic quantum data bus
\cite{Chancellor2012} (see Sec \ref{sub:DB_review} for a brief review).
This data bus takes advantage of antiferromagnetic couplings to transfer
qubits of information adiabatically. First a single qubit, encoded
in a single spin is joined to an even length Heisenberg spin chain
slowly enough such that the adiabatic theorem applies. Then a single
spin on the other end of the chain is separated, again slowly enough
for the adiabatic theorem to apply. As long as the interactions between
the spins on the chain are predominately antiferromagnetic, the qubit
will be successfully transferred from one end of the chain to the
other. This protocol is illustrated in Fig. \ref{fig:transport cartoon}.
Antiferromagnetic spin clusters have been studied for there potential
usefulness in quantum computing in other contexts, for example in
Refs. \cite{Meier2003,Benjamin2003}.

This paper demonstrates how by applying particular unitary operations
to spin chains for single qubit gates, and by using a specific spin
network for a CNOT gate, one can achieve universal holonomic quantum
computation. This method uses open-loop holonomies, meaning that the
Hamiltonian is not necessarily returned to the same state after the
adiabatic process. The methods used here can also be extended to non-adiabatic
implementations of geometrical quantum computing.

Holonomic quantum computation (HQC) was conceived and shown to be
universal by Zanardi and Rasetti \cite{Zanardi1999} and was formulated
in terms of a non-abelian Berry phase. HQC is considered to be an
appealing method for achieving fault tolerant quantum computing because
of its geometrical nature and because it can be implemented adiabatically,
and therefore has all of the advantages of adiabatic quantum computation
\cite{Pachos2001}. Although many implementations of holonomic and
geometric quantum computation are adiabatic, there are examples which
are not \cite{Xu(2012),Mousolou2012}.

Other proposed architectures for holonomic quantum computation use
a variety of architectures, including superconducting systems with
Josephson junctions \cite{Faoro2003}. Further examples propose using
quantum dots \cite{Solinas2003,Mousolou2012}. Single molecule magnets
have been another system of interest \cite{Mousolou2012}. A recent
proposal has also been made for using Holonomies which involve attaching
and removing a spin from a spin 1 chain\cite{Renes2011}. This architecture,
although it looks superficially very similar to ours, performs computations
locally, at the site of the spin rather than by transport as ours
does. Ref. \cite{Renes2011} also proposes an implementation on ultracold
polar molecules. Another interesting proposal involves using a quantum
wire with a twisted cluster state Hamiltonian\cite{Bacon2012}. This
proposal is similar to ours, but implements the twist in a fundamentally
different way.

Most other approaches to HQC involve building a system, and explicitly
calculating the holonomies caused by various manipulations of the
system. In our proposal we start with a process which has a trivial
Berry phase%
\footnote{Here I mean trivial if the adiabatic bus protocol were performed twice,
once to transport the spin to another site and once to return it to
its original location.%
}. Real space twists are then performed on the spin chain used in this
process. Unlike most examples of HQC, all results can be derived without
explicitly considering the curvature of the underlying manifold of
states, all single qubit gates result from the same underlying Hamiltonian
with basis rotations applied to it. 

In principle this idea could be used for more traditional, closed-loop
holonomic quantum computation by twisting a spin chain used to transport
the qubit adiabatically slowly, performing an adiabatic transport
protocol, untwisting the chain adiabatically slowly, and performing
the transport protocol in reverse to return to the initial Hamiltonian.
The net result of this process would be to perform a rotation on the
initial spin which the qubit is encoded and return the spin chain
to its initial state. In this case spin rotation would be achieved
by a non-abelian Berry phase acquired via transport on a closed adiabatic
loop in parameter space, and would therefore be a closed-loop holonomic
gate. Such a process involves unnecessarily many steps however, so
it is more convenient to think of this process in terms of the framework
of open-loop holonomic quantum computation.

The mathematical differences of this approach from others affords
us the advantage that the spectrum of the underlying Hamiltonian is
the same for all twists, meaning that, by construction, all single
qubit gates can be implemented in a way which requires the same annealing
time to reach a given accuracy. This architecture has the advantage
that the only operation it ever requires to be adiabatically performed
is the joining or removal of a spin from a chain or cluster. The nature
of the twists used here also means that a non-adiabatic transport
protocol could be used instead, and universal computation would still
be achieved.

This paper also outlines an implementation of the necessary components
of this design using superconducting flux qubits. Superconducting
flux qubits are a popular architecture for implementing scalable adiabatic
quantum computing \cite{Johnson2011,Harris2010-1,Perdomo2008,vanderPloeg2006,Harris2010-2},
and therefore are a natural choice for designing a scalable holonomic
quantum computer. An additional advantage of the use of superconducting
flux qubits is that the designs tend to have spatially extended qubits
and a high degree of connectivity\cite{Harris2010-2}. The large spatial
extent of the qubits means that a design could be implemented in which
a qubit would only need to be transferred across a small number of
spins to be moved from one location in a computer to any other arbitrary
location. For this reason it is only necessary that the transport
protocol be efficient for short chains, as has already been demonstrated
in \cite{Chancellor2012}, rather than in the thermodynamic limit.

There has been recent experimental work involving quantum annealing
to degenerate ground state manifolds using currently available superconducting
flux qubit hardware\cite{Boixo2012}. In this paper it was demonstrated
experimentally that signatures of quantum behaviors can be observed
in the final state within a degenerate ground state manifold. This
provides an indication that a ground state manifold can be produced
accurately enough on current hardware that quantum effects dominate
over classical effects and design inaccuracies. Although the architecture
proposed here cannot be implemented on the hardware used in \cite{Boixo2012},
this experiment does provide proof of principle for the use of degenerate
manifolds in superconducting flux qubit systems.

While it is not the main focus of this paper, we would also like to
point out that there are other potential methods of implementing this
architecture. One example of such an implementation would be to use
a coupled cavities scheme similar to the one explored in \cite{Chen2010}.
For such an implementation long spin chains may be required, and as
a result properties in the thermodynamic limit may be important. For
such an implementation, the architecture given in this paper could
easily be generalized to a J1-J2 spin chain with $\frac{J_{2}}{J_{1}}\gtrsim0.25$
which is know to be gaped in the thermodynamic limit\cite{Chitra1995}.
In this case, another option would be to implement the architecture
non-adiabatically using the methods described in \cite{Banchi2010,Banchi2011,Apollaro2012,Banchi2011-2}.

\section{\label{part:Single-Q-bit-Gates}Single Q-bit Gates}

\subsection{The Twisted Spin Chain}

Consider initially an antiferromagnetic Heisenberg spin chain. It
has been shown that such a chain can act as a quantum data bus, both
adiabatically\cite{Chancellor2012} and by using the dynamics of its
excitations \cite{Banchi2010,Banchi2011,Apollaro2012,Banchi2011-2}.
The initial Hamiltonian is given by

\begin{equation}
\textrm{H}=\sum_{i=1}^{N-1}\vec{\sigma_{i}}\cdot\vec{\sigma}_{i+1}=\sum_{i=1}^{N-1}(\sigma_{i}^{x}\sigma_{i+1}^{x}+\sigma_{i}^{y}\sigma_{i+1}^{y}+\sigma_{i}^{z}\sigma_{i+1}^{z}).\label{eq:Heisenberg H}
\end{equation}

Now imagine that one inserts a twist into the spin chain by applying
a local unitary transformation of the form $x,y,z\rightarrow x',y',z'$
on $N'=N-L$ spins%
\footnote{Note that this twist can be performed on any bond in the chain, including
one of the bonds, including the bonds which are created or destroyed
in the adiabatic bus protocol. A minimum of three spins is required
for a gate to be implemented adiabatically because of the requirement
of an odd overall chain length by the adiabatic quantum bus protocol.
If a different transport protocol were used, a gate could be implemented
with only 2 spins. %
}, where $x',y',z'$ are all mutually orthogonal to each other. This
yields a new Hamiltonian of the form

\begin{equation}
\textrm{H}_{\textrm{twist}}=\sum_{i=1}^{L-1}\vec{\sigma_{i}}\cdot\vec{\sigma}_{i+1}+\vec{\sigma}_{L}\cdot\vec{\sigma}'_{L+1}+\sum_{j=L+1}^{N-1}\vec{\sigma}'_{j}\cdot\vec{\sigma}'_{j+1}.\label{eq:Htwist}
\end{equation}

\begin{figure}
\includegraphics[scale=0.5]{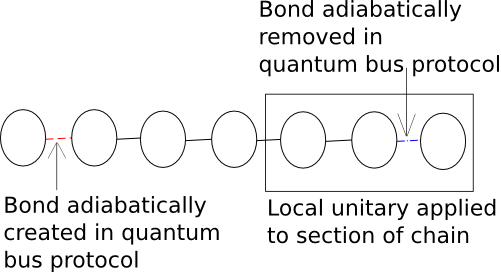}

\caption{\label{fig:twist_protocol}(color online) Illustration of a single
unitary gate implemented by adiabatic transport on a twisted chain.}
\end{figure}

Such a twist does not effect the spectrum of the Hamiltonian, and
therefore the dynamics of the adiabatic quantum bus protocol, or other
quantum bus protocols which may make use of the unitary dynamics of
the Hamiltonian. It is important to note, however, that after transfer
across the chain, the spin will be rotated into the $x',y',z'$ basis.
As we will demonstrate later, transfer through this twisted spin chain
can perform any desired unitary rotation on the qubit being transferred,
and thus can be used to implement any single qubit gate, see Fig.
\ref{fig:twist_protocol}.

One should note that while in this example we consider a simple Heisenberg
spin chain, gates can be implemented in this way on an XYZ spin chain
or a J1-J2 spin chain, or other sufficiently complex quantum spin
Hamiltonians %
\footnote{This argument breaks down in general if one considers cases where
there is not coupling in all three spin directions, for example in
an XY model.%
}. Figuring out which twist to use to perform a given gate can be done
easily and will be illustrated in the next section.

\subsection{Example: Implementing a Hadamard Gate}

The local twist to implement the Hadamard gate is $\sigma^{x}\rightarrow\sigma^{z}$,
$\sigma^{y}\rightarrow-\sigma^{y}$and $\sigma^{z}\rightarrow\sigma^{x}$
. This twist can be calculated without difficulty, for details see
Sec. \ref{sub:Hadamard_calc}.

One can therefore conclude that the Hadamard gate can be implemented
by performing the quantum bus protocol on the Hamiltonian

\begin{equation}
\textrm{H}_{\textrm{Hadamard}}=\sum_{i=1}^{N'-1}\vec{\sigma_{i}}\cdot\vec{\sigma}_{i+1}+\sigma_{N'}^{x}\sigma_{N'+1}^{z}+\label{eq:Hadamard_twist}
\end{equation}

\[
\sigma_{N'}^{z}\sigma_{N'+1}^{x}-\sigma_{N'}^{y}\sigma_{N'+1}^{y}+\sum_{j=N'+1}^{N-1}\vec{\sigma_{j}}\cdot\vec{\sigma}_{j+1}
\]

\subsection{Other Single Qubit Gates}

One can perform similar twists to implement any given single qubit
gate. The calculation to find $x',y',z'$ in Eq. \ref{eq:Htwist}
for other gates can be performed in the same way as the one in the
previous section for the Hadamard. Table \ref{tab:twists} shows how
to implement single qubit gates. These gates are sufficient to perform
an arbitrary unitary operation on a single spin. It is shown in \cite{Nielsen and Chuang}
that any unitary rotation can be approximated to arbitrary precision
with the gates given in table \ref{tab:twists}. We have already shown
how to build an adiabatic quantum bus to move qubit states to arbitrary
locations in the system. Next we discuss that a CNOT gate can be implemented
under this architecture. Then we have demonstrated a universal quantum
computer.

\begin{table}
\begin{tabular}{|c|c|c|c|c|}
\hline 
Gate Name  & Matrix  & $\sigma^{x'}$  & $\sigma^{y'}$  & $\sigma^{z'}$\tabularnewline
\hline 
\hline 
Hadamard  & $\frac{1}{\sqrt{2}}\left(\begin{array}{cc}
1 & 1\\
1 & -1
\end{array}\right)$  & $\sigma^{z}$  & $-\sigma^{y}$  & $\sigma^{x}$\tabularnewline
\hline 
$\frac{\pi}{8}$  & $\left(\begin{array}{cc}
1 & 0\\
0 & \exp(\imath\frac{\pi}{4})
\end{array}\right)$  & $\frac{1}{2}(\sigma^{x}+\sigma^{y}$)  & $\frac{1}{2}(\sigma^{y}-\sigma^{x}$)  & $\sigma^{z}$\tabularnewline
\hline 
phase  & $\left(\begin{array}{cc}
1 & 0\\
0 & \imath
\end{array}\right)$  & $-\sigma^{y}$  & $\sigma^{x}$  & $\sigma^{z}$\tabularnewline
\hline 
NOT%
\footnote{This gate is needed for the construction of the CNOT%
}  & $\left(\begin{array}{cc}
0 & 1\\
1 & 0
\end{array}\right)$  & $\sigma^{x}$  & $-\sigma^{y}$  & $-\sigma^{z}$\tabularnewline
\hline 
\end{tabular}

\caption{ Twists for implementing various single qubit gates\label{tab:twists}}
\end{table}

\section{Implementation of the Controlled NOT Gate}

\subsection{\label{sec:CNOT-design}CNOT design}

\begin{figure}
\includegraphics[scale=0.7]{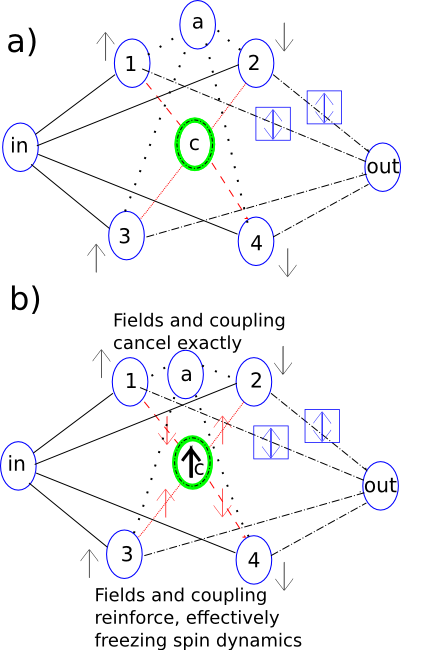}\caption{\label{fig:CNOT_design} (color online) a) Design of a CNOT gate which uses the adiabatic
data bus protocol. Note that one could replace the NOT operation with
any other single q-bit unitary. b) CNOT system with control central
spin up, executes a NOT twist on target spin under quantum bus protocol.
See Tab. \ref{tab:CNOT_legend} for the meaning of various symbols.
Labels are based on Eq.\ref{eq:cnot hamiltonian}.}
\end{figure}

Let us now turn our attention to the implementation of a controlled
NOT (CNOT) using an adiabatic quantum bus protocol. In Fig. \ref{fig:CNOT_design}a)
we show a design for such a gate. The time dependent Hamiltonian for
this gate is

\begin{multline}
\textrm{H}_{CNOT}(t;h,t_{fin})=\lambda(t;t_{fin})\vec{\sigma}_{in}\cdot(\vec{\sigma}_{1}+\vec{\sigma}_{2}+\vec{\sigma}_{3}+\vec{\sigma}_{4})\\
+\vec{\sigma}_{a}\cdot(\vec{\sigma}_{1}+\vec{\sigma}_{2}+\vec{\sigma}_{3}+\vec{\sigma}_{4})+h((\sigma_{1}^{z}-\sigma_{2}^{z})(1-\sigma_{c}^{z})+(\sigma_{3}^{z}-\sigma_{4}^{z})(1+\sigma_{c}^{z}))\\
+(1-\lambda(t;t_{fin}))(\vec{\sigma}'_{out}\cdot(\vec{\sigma}_{1}+\vec{\sigma}_{2}))+\vec{\sigma}{}_{out}(\vec{\sigma}_{3}+\vec{\sigma}_{4}),\label{eq:cnot hamiltonian}
\end{multline}

\[
\lambda(t;t_{fin})=\begin{cases}
0 & t<0\\
\frac{t}{t_{fin}} & 0\leq t\leq t_{fin}\\
1 & t>t_{fin}
\end{cases}.
\]

Here ``in'' refers to the spin which is the input spin, where the
target qubit is initially encoded; ``out'' refers to the spin to
which the target qubit is transferred to, ``c'' refers to the control
qubit, ``a'' to an ancilla to make the number of intermediate spins
odd. The other 4 spins are assigned numbers 1-4. $\vec{\sigma}'_{out}$
refers to a NOT twist being performed on these Pauli matrices, see
Tab. \ref{tab:twists}. This gate operates by having 2 channels through
which a qubit of information can pass. One channel, consisting of
spins 3 and 4, allows the information to pass through the gate unaltered,
while another channel, consisting of spins 1 and 2 performs a twist
on the qubit as it travels though the gate. The control spin c controls
though which channel the information travels. The control spin is
connected with Ising type coupling to spins 1-4 in such a way that
when the control spin is up the external field on spins 1 and 2 cancels
with the effect of the Ising bond with spin c because $(\frac{1}{2}-<\sigma_{c}^{z}>)=0$,
and the information can easily pass though these spins. On the other
hand $(\frac{1}{2}+<\sigma_{c}^{z}>)=1$. So spins 3 and 4 both have
an effective magnetic field of 2h. For sufficiently large h these
spins are frozen in the direction of the field and will therefore
not be able to transport any information. As we show in Fig. \ref{fig:CNOT_design}b)
the net effect is that information all travels though spins 1 and
2, and therefore a NOT twist is performed. In the case where the spin
c is in the down direction, information will instead be allowed to
travel though spins 3 and 4 and blocked on spins 1 and 2. Therefore
in that case the gate acts trivially on the qubit.

Any state of the control spin c can be expressed as $\ket{\psi_{c}}=a\ket{\uparrow}+b\ket{\downarrow}$
where a and b are complex numbers. With an arbitrary input state $\ket{\psi_{in}}=\alpha\ket{\uparrow}+\beta\ket{\downarrow}$
we have before the gate $\ket{\psi_{init}}=\ket{\psi_{in}}\otimes\ket{\psi_{c}}=\alpha a\ket{\uparrow\uparrow}+\alpha b\ket{\uparrow\downarrow}+\beta a\ket{\downarrow\uparrow}+\beta b\ket{\downarrow\downarrow}$.
After the gate is performed, the final state becomes $\ket{\psi_{fin}}=\alpha a\ket{\downarrow\uparrow}+\alpha b\ket{\uparrow\downarrow}+\beta a\ket{\uparrow\uparrow}+\beta b\ket{\downarrow\downarrow}$.
From these general states we see that

\begin{equation}
\ket{\psi_{fin}}=\left(\begin{array}{cccc}
1 & 0 & 0 & 0\\
0 & 1 & 0 & 0\\
0 & 0 & 0 & 1\\
0 & 0 & 1 & 0
\end{array}\right)\ket{\psi_{init}},
\end{equation}

which is the definition of a controlled NOT gate \cite{Nielsen and Chuang}.

\begin{table}
\begin{tabular}{|c|c|}
\hline 
Symbol  & Meaning\tabularnewline
\hline 
\hline 
\includegraphics[scale=0.7]{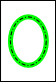}  & spin $\frac{1}{2}$ control spin\tabularnewline
\hline 
\includegraphics[scale=0.7]{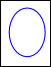}  & spin $\frac{1}{2}$ working spin\tabularnewline
\hline 
\includegraphics[scale=0.7]{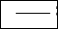}  & initial AF Heisenberg bond (weak)\tabularnewline
\hline 
\includegraphics[scale=0.7]{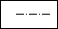}  & final AF Heisenberg bond (weak)\tabularnewline
\hline 
\includegraphics[scale=0.7]{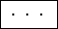}  & fixed AF Heisenberg bond (weak)\tabularnewline
\hline 
\includegraphics[scale=0.7]{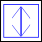}  & twist to perform NOT operation\tabularnewline
\hline 
\includegraphics[scale=0.7]{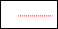}  & fixed F Ising bond (strong)\tabularnewline
\hline 
\includegraphics[scale=0.7]{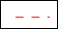}  & fixed AF Ising bond (strong)\tabularnewline
\hline 
\includegraphics[scale=0.7]{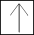}  & fixed ``up'' magnetic field (strong)\tabularnewline
\hline 
\includegraphics[scale=0.7]{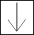}  & fixed ``down'' magnetic field (strong)\tabularnewline
\hline 
\end{tabular}

\caption{\label{tab:CNOT_legend}Legend of symbols used in Figs. \ref{fig:CNOT_design}.}
\end{table}

\subsection{Performance of Controlled NOT Gate}

We need to test how this design for a CNOT gate performs because we
cannot rely on previous work to show that the qubit is actually transferred
accurately. The two free parameters in Eq. \ref{eq:cnot hamiltonian}
are the strength of the Ising bonds and the fields which we denote
by h, and the time for the protocol to be performed, $t_{fin}.$

\begin{figure}
\includegraphics[scale=0.5]{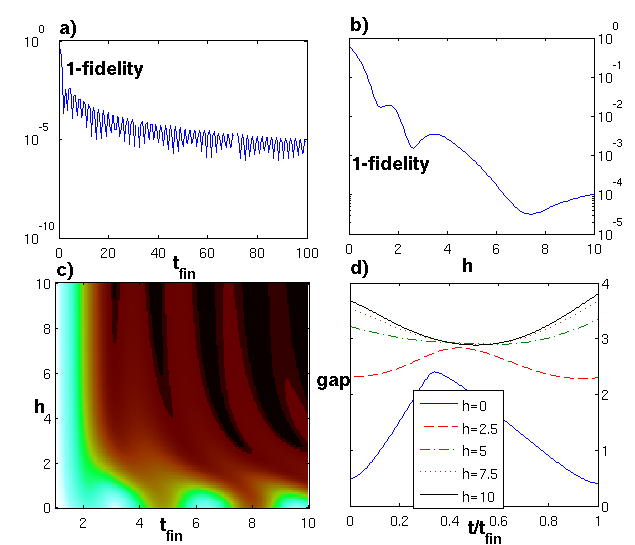}

\caption{\label{fig:CNOT_performance} (color online) Measures of performance
of the CNOT gate a) one minus fidelity of the output spin versus $t_{fin}$
for h=10 b) one minus fidelity of the output spin versus h for $t_{fin}$=10
c) output fidelity for initial up spin with a NOT performed versus
h and $t_{fin}$ light is larger (more positive), dark is smaller
(more negative) d) gap versus $\frac{t}{t_{fin}}$ for various values
of h. The field h and the gap are given in units of the Heisenberg
bond energy.}
\end{figure}

We now examine whether this Hamiltonian actually implements a CNOT
gate effectively for reasonable values of h and $t_{fin}$. To test
this we need to answer 2 questions. Firstly, is the system close enough
to the adiabatic limit for reasonable values of $t_{fin}$? Secondly,
is the desired effect of shutting off one possible path for the information
achieved for reasonable values of h? To answer these questions, we
examine the overlap of the final output state (final state of the
``out'' qubit in Fig. \ref{fig:CNOT_design}) with the expected
output state from a controlled NOT gate (Fig.\ref{fig:CNOT_performance}).
Note that because the case where a NOT gate is performed, and the
case where the gate acts trivially are related by a simple unitary
transformation on the Hamiltonian, acceptable performance in one of
these cases implies acceptable performance in the other. Averaging
over different initial states is therefore unnecessary as it would
yield the exact same result as any particular choice of control and
input states.

Fig. \ref{fig:CNOT_performance} demonstrates the effectiveness of
this gate. Fig. \ref{fig:CNOT_performance} a) shows that for a moderate
field and Ising bond strength the gate can be made to perform well
as long as the annealing time is sufficient. This figure also shows
that the gate can continually be made more effective by running it
longer without having to raise h. The oscillations in the fidelity
are related to the time scale of small excitations produced during
the annealing process. In many applications one may have enough control
over $t_{fin}$that the annealing time can be chosen in a way that
the process lies near one of the local minima of error shown in Fig.
\ref{fig:CNOT_performance} a).

Fig. \ref{fig:CNOT_performance} b) shows the effect of h on fidelity
for an annealing time of 10 (in units of inverse Heisenberg couplings).
In this figure one can see that increasing h is ineffective at improving
performance above a certain value. This indicates that at this point
the field is already effectively completely blocking one path that
the information transport can take. Fig. \ref{fig:CNOT_performance}
c) shows the combined effects of h and $t_{fin}$on the output polarization.
It is consistent with the conclusions we have reached from a) and
b). Finally, Fig. \ref{fig:CNOT_performance} d) shows that the system
gap remains quite large throughout the process. It also demonstrates
that beyond a certain value of h the gap does not increase significantly
with changing h, which is consistent with the picture of one path
being completely closed to information transfer. It is interesting
to note that increasing h increases the gap throughout the process.

\section{Implementation Using Superconducting Flux Qubits}

To build an implementation of this holonomic architecture based on
superconducting flux qubits one needs to design circuits that implement
Heisenberg spins and the appropriate gate couplings between spins.
Fortunately significant work has already been done, for example in
Refs. \cite{Johnson2011,Harris2010-1,Perdomo2008,vanderPloeg2006,Harris2010-2},
towards the design of couplers for superconducting flux qubits. However
since the previously discussed schemes were based on Ising spin systems,
we still need to establish a method for designing circuits which emulate
Heisenberg spins. It is interesting to point out that this computational
architecture works with the limited connectivity of the designs proposed
in Refs. \cite{Johnson2011,Harris2010-1,Perdomo2008,vanderPloeg2006,Harris2010-2}.
Specifically the CNOT gate we have proposed fits in a single 2x4 chimera
lattice cell like the one used in Ref. \cite{Harris2010-2}.

\subsection{Flux Qubit Motivation:}

It has been shown in \cite{Harris2010-1} that a qubit which behaves
like an Ising spin can be constructed from a Hamiltonian of the form

\begin{equation}
\textrm{H}_{Ising}=\sum_{n=1}^{2}(\frac{Q_{n}^{2}}{2C_{n}}+U_{n}\frac{(\phi_{n}-\phi_{n}^{x})^{2}}{2})-U_{q}\cos(\alpha_{1}\phi_{1})\cos(\alpha_{2}\phi_{2}),\label{eq:Ising_Ham}
\end{equation}

where the applied fluxes $\phi_{n}^{x}$ act as effective magnetic
fields, altering the shape of a potential well for the system in a
way that mimics a spin constrained to move in a plane. The constants
$\alpha$ simply act to scale the effect of the flux. To construct
a Heisenberg qubit one needs to add a third direction, leading to
a Hamiltonian of the form

\begin{equation}
\textrm{H}_{Heis.}=\sum_{n=1}^{3}(\frac{Q_{n}^{2}}{2C_{n}}+U_{n}\frac{(\phi_{n}-\phi_{n}^{x})^{2}}{2})\label{eq:Heisenberg_Ham}
\end{equation}

\[
-U_{q}\cos(\alpha_{1}\phi_{1})\cos(\alpha_{2}\phi_{2})\cos(\alpha_{3}\phi_{3}).
\]

Such a Hamiltonian would allow the shape of the potential well to
be changed along 3 directions and would therefore mimic a Heisenberg
spin rather than an Ising spin. Previously proposed designs also only
couple qubits along one direction. If a new type of coupler were added
to the currently implemented circuits which couple the spins in the
y direction in addition the z direction, then an XY model could be
implemented. To implement an XYZ Heisenberg model, one needs to both
design qubits which are not constrained to lie in a plane and build
3 types of couplers, one for each direction in space.

\section{Flux Qubit Design:}

First consider a CCJJ (Compound-Compound Josephson Junction) circuit
as defined in \cite{Harris2010-1}.

The effective Hamiltonian of this circuit is given by \cite{Harris2010-1}

\begin{equation}
\textrm{H}=\sum_{n}(\frac{Q_{n}^{2}}{2C_{n}}+U_{n}\frac{(\phi_{n}-\phi_{n}^{x})^{2}}{2})-U_{q}\beta_{eff}\cos(\phi_{q}-\phi_{q}^{0}),\label{eq:Ham}
\end{equation}

where $n\in\{q,cjj,l,r\}$. Note that there are more indeces than
in \cite{Harris2010-1} because we do not have the condition that
$\phi_{L}=\phi_{L}^{x}$ or $\phi_{R}=\phi_{R}^{x}$. The first two
terms of the Hamiltonian in Eq.\ref{eq:Ham} are not important for
what we are trying to demonstrate here \cite{Harris2010-1}. The definition
of all of these terms can be found in Eq. B4b-f in \cite{Harris2010-1},
and in Sec. \ref{sub:qubit_asssumptions} of this paper.

\begin{figure}
\includegraphics[scale=0.4]{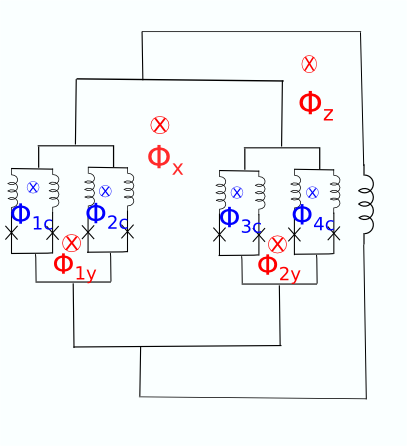}

\caption{(color online) One can build a CCCJJ by replacing every Josephson
Junction with a pair of parallel junctions in the CCJJ. By controlling
the flux in any of the smallest loops ($\Phi_{nc}$) one can effectively
change the critical current of the junction pair and compensate for
manufacturing errors. A similar example with a CJJ and a CCJJ can
be found in \cite{Harris2010-1}\label{fig:CCCJJ}. }
\end{figure}

Let us make the simplifying assumption that all of the critical currents
are equal for all junctions. In practice there is variability in junction
fabrication, but this error can be compensated by building a CCCJJ
device (see Fig. \ref{fig:CCCJJ}). Let us also assume that our circuit
is designed in such a way that we can inductively couple the left
and right loop to each other very strongly such that $\phi_{y}\equiv\phi_{L}=\phi_{R}$
and $\phi_{y}^{x}\equiv\phi_{L}^{x}=\phi_{R}^{x}$. These assumptions
cause the equations to simplify greatly (see Sec. \ref{sub:qubit_asssumptions}),
yielding

\begin{equation}
\beta_{eff}=\beta_{+}\cos(\frac{\phi_{ccjj}}{2}),\label{eq:Beffsimp}
\end{equation}

\[
\beta_{+}\equiv2\beta_{L}=2\beta_{R},
\]

\begin{equation}
\beta_{L(R)}=\frac{4\pi L_{q}I_{c}}{\Phi_{0}}\cos(\frac{\phi_{y}}{2}).\label{eq:Blrsimp}
\end{equation}

This leads to an effective Hamiltonian of the form

\begin{equation}
\textrm{H}=\sum_{n}(\frac{Q_{n}^{2}}{2C_{n}}+U_{n}\frac{(\phi_{n}-\phi_{n}^{x})^{2}}{2})-U_{q}\beta_{+}\cos(\frac{\phi_{ccjj}}{2})\cos(\phi_{q})\label{eq:Hsimp}
\end{equation}

\begin{figure}
\includegraphics{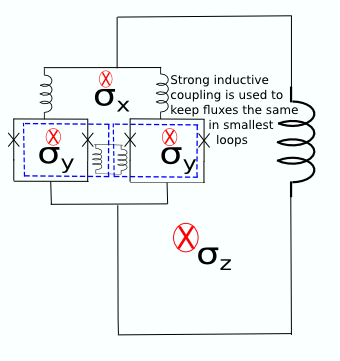}

\caption{(color online) Design using strong inductive coupling between small
loops, where fluxes mimic various magnetic fields applied to a spin.
Note that this design assumes that all Josephson junctions are identical.\label{fig:Circ_ind}}
\end{figure}

When we substitute in $\beta_{+}$from Eq. \ref{eq:Blrsimp}, this
Hamiltonian becomes

\begin{equation}
\textrm{H}=\sum_{n}(\frac{Q_{n}^{2}}{2C_{n}}+U_{n}\frac{(\phi_{n}-\phi_{n}^{x})^{2}}{2})\label{eq:Hfin}
\end{equation}

\[
-U_{q}\frac{8\pi L_{q}I_{c}}{\Phi_{0}}\cos(\frac{\phi_{y}}{2})\cos(\frac{\phi_{x}}{2})\cos(\phi_{z}),
\]

which is of the form given in Eq. \ref{eq:Heisenberg_Ham}. The corresponding
circuit is shown in Fig. \ref{fig:Circ_ind}.

\section{Conclusions}

We have demonstrated an architecture for a universal quantum computer
using Heisenberg spin chains and clusters. This architecture has the
advantage that it can be implemented adiabatically and therefore has
all of the advantages of adiabatic quantum computing. It has already
been demonstrated in \cite{Chancellor2012} that the single qubit
gates and data bus used in this computer can be implemented with high
fidelity for reasonable annealing time. We further demonstrate that
a controlled NOT gate can be implemented at high fidelity with a reasonably
short annealing time and reasonable Hamiltonian parameters.

In addition to suggesting an architecture, we also propose a design
to physically realize this architecture. We suggest a method for building
superconducting flux qubit systems which model the low energy degrees
of freedom of a Heisenberg model. Because the architecture we propose
can be implemented adiabatically, only the low energy degrees of freedom
need to be reproduced. We choose a superconducting flux qubit implementation
because of the experimental success of these systems in performing
non-universal adiabatic quantum computing with an Ising spin glass
model. Furthermore it has been shown that these Ising spin glass models
can be realized accurately enough that the degeneracy of the ground
state manifolds is not broken. The universal Heisenberg spin based
computer would represent a significant improvement over the current
non-universal Ising systems because it would allow these computers
to implement important algorithms which the Ising spin glass system
has not been able to, such as Shors algorithm for factoring large
numbers \cite{Johnson2011,Harris2010-1,Perdomo2008,vanderPloeg2006,Harris2010-2}. 
\begin{acknowledgments}
The authors would like to thank P. Zanardi, I. Marvian, S. Boixo,
D. A. Lidar, L. C. Venuti, and S. Takehashi for helpful conversations.
Some numerical calculations were performed on the USC high performance
computing cluster. This research is partially supported by the ARO
MURI grant W911NF-11-1-0268. 
\end{acknowledgments}

\section{Appendix\label{sec:Appendix}}

\subsection{Review of adiabatic quantum data bus \label{sub:DB_review}}

The adiabatic data bus protocol discussed here is a process in which
an even length anti-ferromagnetic Heisenberg spin chain prepared in
its ground state is joined adiabatically slowly to a single spin in
a state $\ket{\psi}$. A spin is also removed from the opposite side
of the chain, again adiabatically slowly. The final removed spin will
again be in the state $\ket{\psi}$. The Hamiltonian for this protocol
is

\begin{equation}
H_{AQB}(N,t_{fin})=A(t,t_{fin})\vec{\sigma_{1}}\vec{\sigma}_{2}+\sum_{i=2}^{N-2}\vec{\sigma_{i}}\vec{\sigma}_{i+1}\label{eq:qbus}
\end{equation}

\[
+B(t,t_{fin})\vec{\sigma_{N-1}}\vec{\sigma}_{N},
\]

where $A(t,t_{fin})$ and $B(t,t_{fin})$ are the annealing schedules,
functions chosen such that $A(t\leq0,t_{fin})=0$, $A(t\geq t_{fin},t_{fin})=1$,
$B(t\leq0,t_{fin})=1$, and $B(t\geq t_{fin},t_{fin})=0$. This protocol
can also be generalized to an XYZ and J1-J2 spin chain (see \cite{Chancellor2012}
for more details). For simplicity let us consider this protocol with
a linear annealing schedule such that 

\begin{equation}
A(t,t_{fin})=\begin{cases}
0 & t\leq0\\
\frac{t}{t_{fin}} & 0<t<t_{fin}\\
1 & t\geq t_{fin}
\end{cases},\label{eq:annealing schedule}
\end{equation}

and 
\[
B(t,t_{fin})=1-A(t,t_{fin}).
\]

\begin{figure}
\includegraphics[scale=0.4]{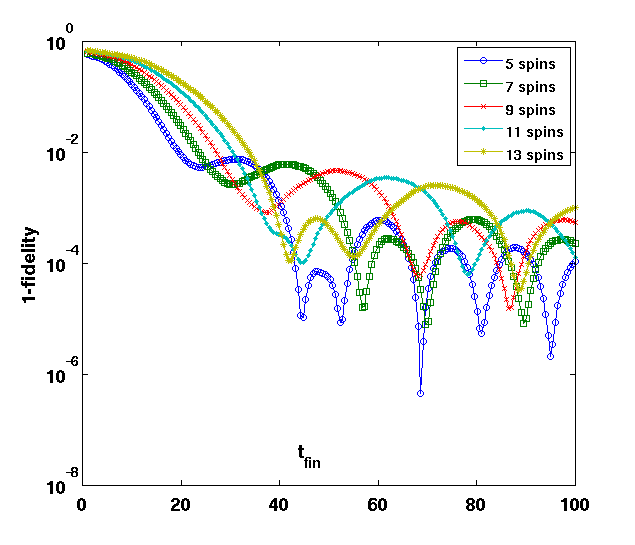}\caption{\label{fig:qbus fid v t}(color online)One minus fidelity of qubit
transport versus annealing time for quantum bus protocol with linear
annealing schedule. t is in units of the Heisenberg bond energy.}

\end{figure}

The results of this annealing protocol can be seen in Fig. \ref{fig:qbus fid v t}.

\subsection{Calculation of twist for Hadamard gate\label{sub:Hadamard_calc}}

Showing how to implement any given single spin gate using this method
is straightforward. Take for example the Hadamard gate $\mathcal{H}$,

\begin{equation}
\mathcal{H}\ket{\psi}=\frac{1}{\sqrt{2}}\left(\begin{array}{cc}
1 & 1\\
1 & -1
\end{array}\right)\ket{\psi}.\label{eq:Hadamard}
\end{equation}

We now consider the action of $\mathcal{H}$ on the eigenvectors of
the Pauli spin matrices, first for $\sigma^{x}$:

\[
x_{+}=\frac{1}{\sqrt{2}}\left(\begin{array}{c}
1\\
1
\end{array}\right)\rightarrow x'_{+}=\mathcal{H}x_{+}=\left(\begin{array}{c}
1\\
0
\end{array}\right)=z_{+}
\]

\[
x_{-}=\frac{1}{\sqrt{2}}\left(\begin{array}{c}
1\\
-1
\end{array}\right)\rightarrow x'_{-}=\mathcal{H}x_{-}=\left(\begin{array}{c}
0\\
1
\end{array}\right)=z_{-}
\]

Similarly for $\sigma^{z}$:

\[
z_{+}=\left(\begin{array}{c}
1\\
0
\end{array}\right)\rightarrow z'_{+}=\mathcal{H}z_{+}=\frac{1}{\sqrt{2}}\left(\begin{array}{c}
1\\
1
\end{array}\right)=x_{+}
\]

\[
z_{-}=\left(\begin{array}{c}
0\\
1
\end{array}\right)\rightarrow z'_{-}=\mathcal{H}z_{-}=\frac{1}{\sqrt{2}}\left(\begin{array}{c}
1\\
-1
\end{array}\right)=x_{-}
\]

And for $\sigma^{y}$:

\[
y_{+}=\frac{1}{\sqrt{2}}\left(\begin{array}{c}
1\\
\imath
\end{array}\right)\rightarrow y'_{+}=\mathcal{H}y_{+}
\]

\[
=\frac{1}{2}\left(\begin{array}{c}
1+\imath\\
\imath-1
\end{array}\right)=\frac{\imath+1}{\sqrt{2}}\left(\begin{array}{c}
1\\
-\imath
\end{array}\right)=y_{-}\exp(\imath\phi)
\]

\[
y_{-}=\frac{1}{\sqrt{2}}\left(\begin{array}{c}
1\\
-\imath
\end{array}\right)\rightarrow y'_{-}=\mathcal{H}y_{-}
\]

\[
=\frac{1}{2}\left(\begin{array}{c}
1+\imath\\
\imath-1
\end{array}\right)=\frac{\imath-1}{\sqrt{2}}\left(\begin{array}{c}
1\\
\imath
\end{array}\right)=y_{+}\exp(-\imath\phi)
\]

In this case the phase factor ($\phi=\frac{\pi}{4}$) is irrelevant
because of the overall U(1) symmetry.

\subsection{Detailed discussion of simplifying assumptions for superconducting
flux qubits\label{sub:qubit_asssumptions}}

The last term $U_{q}$ in Eq. \ref{eq:Ham} is a constant which is
not relevant for this discussion. However the other constants in this
term are relevant and are defined as follow (Eq. B4b-f in \cite{Harris2010-1})

\begin{equation}
\beta_{eff}=\beta_{+}\cos(\frac{\gamma}{2})\sqrt{1+(\frac{\beta_{-}}{\beta_{+}}\tan(\frac{\gamma}{2}))^{2}},\label{eq:beff}
\end{equation}

\begin{equation}
\phi_{q}^{0}=\frac{\phi_{L}^{0}+\phi_{R}^{0}}{2}+\gamma_{0},\label{eq:f0q}
\end{equation}

\begin{equation}
\gamma\equiv\phi_{ccjj}-(\phi_{L}^{0}-\phi_{R}^{0}),\label{eq:gamma}
\end{equation}

\begin{equation}
\gamma_{0}\equiv-\arctan(\frac{\beta_{-}}{\beta_{+}}\tan(\frac{\gamma}{2})),\label{eq:gamma0}
\end{equation}

\begin{equation}
\beta_{\pm}\equiv\beta_{L}\pm\beta_{R}.\label{eq:Bpm}
\end{equation}

Here we need the additional definitions:

\begin{equation}
\beta_{L(R)}=\beta_{L(R),+}\cos(\frac{\phi_{L(R)}}{2})\sqrt{1+(\frac{\beta_{L(R),-}}{\beta_{L(R),+}}\tan(\frac{\phi_{L(R)}}{2}))^{2}},\label{eq:Blr}
\end{equation}

\begin{equation}
\phi_{L(R)}^{0}=\arctan(\frac{\beta_{L(R),-}}{\beta_{L(R),+}}\tan(\frac{\phi_{L(R)}}{2})),\label{eq:f0lr}
\end{equation}

\begin{equation}
\beta_{L(R),\pm}=\frac{2\pi L_{q}(I_{1(3)}\pm I_{2(4)})}{\Phi_{0}}.\label{eq:Blrpm}
\end{equation}

Let us make the simplifying assumption that all of the critical currents
are equal, $I_{1}=I_{2}=I_{3}=I_{4}$. In practice there is variability
in junction fabrication, but this error can be compensated by building
a CCCJJ device (see Fig. \ref{fig:CCCJJ}). This assumption causes
the equations to simplify greatly because $\beta_{L(R),-}\rightarrow0$,
which has the consequence that $\phi_{L(R)}^{0}\rightarrow0$ and
$\gamma\rightarrow\phi_{ccjj}$. We can further assume that in our
design that $\phi_{L}=\phi_{R}$, this additional assumption causes
$\beta_{-}=0$ and $\gamma_{0}\rightarrow0$, which in turn causes
$\phi_{q}^{0}\rightarrow0$. After this simplification we now have

\begin{equation}
\beta_{eff}=\beta_{+}\cos(\frac{\phi_{ccjj}}{2}),\label{eq:Beffsimp-1}
\end{equation}

\begin{equation}
\beta_{L(R)}=\frac{4\pi L_{q}I_{c}}{\Phi_{0}}\cos(\frac{\phi_{L(R)}}{2}).\label{eq:Blrsimp-1}
\end{equation}

\end{document}